\documentclass[12pt,titlepage]{article}
\usepackage{amsmath}
\usepackage{amssymb}
\usepackage[dvips]{graphicx}
\usepackage{amsfonts}
\usepackage[portuguese]{babel}
\usepackage[font=small,format=plain,labelfont=bf,up,textfont=it,up]{caption}

\begin{document}

\title{\textbf{O oscilador harm\^{o}nico singular revisitado\thanks{To appear in Revista Brasileira de Ensino de F\'{\i}sica}}\\
{\small The singular harmonic oscillator revisited}}
\date{}

\author{Douglas R. M. Pimentel e Antonio S. de Castro\thanks{
Enviar correspond\^{e}ncia para A. S. de Castro. E-mail: castro@pq.cnpq.br.%
} \\
\\
UNESP - Campus de Guaratinguet\'{a}\\
Departamento de F\'{\i}sica e Qu\'{\i}mica, \\
Universidade Estadual Paulista \textquotedblleft J\'{u}lio de Mesquita
Filho\textquotedblright, \\
Guaratinguet\'{a}, SP, Brasil}

\maketitle

\begin{abstract}
Investiga-se a equa\c{c}\~{a}o de Schr\"{o}dinger unidimensional com o
oscilador harm\^{o}nico singular. A hermiticidade dos operadores associados
com quantidades f\'{\i}sicas observ\'{a}veis \'{e} usada como crit\'{e}rio
para mostrar que o oscilador singular atrativo ou repulsivo exibe um n\'{u}%
mero infinito de solu\c{c}\~{o}es aceit\'{a}veis, contanto que o par\^{a}%
metro respons\'{a}vel pela singularidade seja maior que um certo valor cr%
\'{\i}tico, em discord\^{a}ncia com a literatura. O problema definido em
todo o eixo exibe dupla degeneresc\^{e}ncia no caso do oscilador singular e
intrus\~{a}o de adicionais n\'{\i}veis de energia no caso do oscilador n\~{a}%
o-singular. Outrossim, mostra-se que a solu\c{c}\~{a}o do oscilador singular
n\~{a}o pode ser obtida a partir da solu\c{c}\~{a}o do oscilador n\~{a}%
o-singular via teoria da perturba\c{c}\~{a}o.

\bigskip

\noindent \textbf{Palavras-chave:} oscilador harm\^{o}nico, potencial
singular, degeneresc\^{e}ncia, colapso para o centro \newline

\bigskip

\bigskip

\bigskip

\noindent {\small {The one-dimensional Schr\"{o}dinger equation with the
singular harmonic oscillator is investigated. The Hermiticity of the
operators related to observable physical quantities is used as a criterion
to show that the attractive or repulsive singular oscillator exhibits an
infinite number of acceptable solutions provided the parameter responsible
for the singularity is greater than a certain critical value, in
disagreement with the literature. The problem for the whole line exhibits a
two-fold degeneracy in the case of the singular oscillator, and the
intrusion of additional solutions in the case of a nonsingular oscillator.
Additionally, it is shown that the solution of the singular oscillator can
not be obtained from the nonsingular oscillator via perturbation theory.}}

\bigskip

\end{abstract}

\section{Introdu\c{c}\~{a}o}

O oscilador harm\^{o}nico \'{e} um dos mais importantes sistemas em mec\^{a}%
nica qu\^{a}ntica porque ele apresenta solu\c{c}\~{a}o em forma fechada e
isto pode ser \'{u}til para gerar solu\c{c}\~{o}es aproximadas ou solu\c{c}%
\~{o}es exatas para v\'{a}rios problemas. O oscilador harm\^{o}nico \'{e}
costumeiramente resolvido com o m\'{e}todo de solu\c{c}\~{a}o em s\'{e}ries
de pot\^{e}ncias (veja, e.g., \cite{som}) e o m\'{e}todo alg\'{e}brico
(veja, e.g., \cite{sak}), e tamb\'{e}m por meio de t\'{e}cnicas de integra%
\c{c}\~{a}o de trajet\'{o}ria (veja, e.g., \cite{fey}). Recentemente, o
oscilador harm\^{o}nico unidimensional foi abordado com os m\'{e}todos
operacionais da transformada de Fourier \cite{mun} e da transformada de
Laplace \cite{pim}.

A equa\c{c}\~{a}o de Schr\"{o}dinger com um potencial quadr\'{a}tico
acrescido de um termo inversamente quadr\'{a}tico, conhecido como oscilador
harm\^{o}nico singular, tamb\'{e}m \'{e} um problema exatamente sol\'{u}vel
\cite{lan}-\cite{pat}. A bem da verdade, o problema geral de espalhamento e
estados ligados em potenciais singulares \'{e} um tema antigo (veja, e.g.,
\cite{cas}). O caso do oscilador harm\^{o}nico singular com par\^{a}metros
do potencial dependentes do tempo t\^{e}m sido alvo de investiga\c{c}\~{a}o
recente \cite{cam}. \ O oscilador singular se presta para a constru\c{c}\~{a}%
o de modelos sol\'{u}veis de $N$ corpos \cite{cal1}, tanto quanto como base
para expans\~{o}es perturbativas e an\'{a}lise variacional para osciladores
harm\^{o}nicos acrescidos de termos com singularidades muito mais fortes que
o termo inversamente quadr\'{a}tico \cite{hal1}. O oscilador singular tamb%
\'{e}m tem sido utilizado em mec\^{a}nica qu\^{a}ntica relativ\'{\i}stica
\cite{nag}. A exata solubilidade do oscilador singular pode ser constatada
nas refer\^{e}ncias \cite{lan} e \cite{cm} para o caso tridimensional, e tamb%
\'{e}m \'{e} patente nas refer\^{e}ncias \cite{gol} e \cite{ha} para o caso
unidimensional restrito ao semieixo positivo. Duas refer\^{e}ncias mais
recentes abordam o problema unidimensional em todo o eixo \cite{bg}-\cite{pr}%
. Na Ref. \cite{bg} n\~{a}o h\'{a} detalhes da solu\c{c}\~{a}o do problema
nem men\c{c}\~{a}o \`{a}s poss\'{\i}veis degeneresc\^{e}ncias, e l\'{a}
consta que, para um oscilador singular atrativo, a part\'{\i}cula colapsa
para o ponto $x=0$ (\textit{The particle collapses to the point $x=0$}). Na
Ref. \cite{pr}, Palma e Raff esmiu\c{c}am o problema com o potencial
singular repulsivo, concluem apropriadamente sobre a degeneresc\^{e}ncia e
simplesmente afirmam que n\~{a}o h\'{a} estado fundamental no caso do
oscilador singular atrativo (\textit{the attractive potential has no lower
energy bound}).

O intuito deste trabalho \'{e} perscrutar a equa\c{c}\~{a}o de Schr\"{o}%
dinger unidimensional com o oscilador harm\^{o}nico singular. Veremos que, al%
\'{e}m de uma cr\'{\i}tica \`{a} literatura concernente a um problema de
interesse recente e j\'{a} cristalizado em livros-texto largamente
conhecidos, a abordagem dos estados ligados do oscilador harm\^{o}nico
singular que se presencia neste trabalho permite aos estudantes de mec\^{a}%
nica qu\^{a}ntica e f\'{\i}sica matem\'{a}tica dos cursos de gradua\c{c}\~{a}%
o em f\'{\i}sica o contato com equa\c{c}\~{o}es diferenciais singulares e o
comportamento assint\'{o}tico de suas solu\c{c}\~{o}es, fun\c{c}\~{a}o
hipergeom\'{e}trica confluente, polin\^{o}mios de Laguerre e de Hermite e
outras fun\c{c}\~{o}es especiais, valor principal de Cauchy de integrais impr%
\'{o}prias, condi\c{c}\~{a}o de hermiticidade sobre operadores associados
com grandezas f\'{\i}sicas observ\'{a}veis e o descarte de solu\c{c}\~{o}es
esp\'{u}rias, condi\c{c}\~{o}es de contorno e analiticidade das solu\c{c}%
\~{o}es na vizinhan\c{c}a de pontos singulares, paridade e extens\~{o}es sim%
\'{e}tricas e antissim\'{e}tricas de autofun\c{c}\~{o}es, degeneresc\^{e}%
ncia em sistemas unidimensionais, transi\c{c}\~{a}o de fase e surgimento de n%
\'{\i}veis intrusos, \textit{et cetera}. Seguramente, a profus\~{a}o de
conceitos e t\'{e}cnicas \'{e} do interesse de estudantes e instrutores. Com
o crit\'{e}rio de hermiticidade dos operadores associados com quantidades f%
\'{\i}sicas observ\'{a}veis, o tratamento do problema p\~{o}e \`{a} vista
dos leitores que o oscilador singular, seja atrativo ou repulsivo, exibe um n%
\'{u}mero infinito de solu\c{c}\~{o}es aceit\'{a}veis desde que o par\^{a}%
metro respons\'{a}vel pela singularidade seja maior que um certo valor cr%
\'{\i}tico. Veremos que o espectro de energia \'{e} uma fun\c{c}\~{a}o mon%
\'{o}tona do par\^{a}metro respons\'{a}vel pela singularidade e que a
energia do estado fundamental do oscilador singular, independentemente do
sinal de tal par\^{a}metro, \'{e} sempre maior que dois ter\c{c}os da
energia do estado fundamental do oscilador n\~{a}o-singular para o problema
definido no semieixo, e sempre maior que o dobro da energia do estado
fundamental do oscilador n\~{a}o-singular para o problema definido em todo o
eixo. Mostramos que o problema definido em todo o eixo \ nos conduz \`{a}
dupla degeneresc\^{e}ncia no caso do potencial singular e \`{a} intrus\~{a}o
de adicionais n\'{\i}veis de energia no caso do oscilador n\~{a}o-singular
(relacionados com as autofun\c{c}\~{o}es de paridade par). O robusto crit%
\'{e}rio de hermiticidade do operador associado com a energia cin\'{e}tica
(ou potencial) mostra-se suficiente para descartar solu\c{c}\~{o}es ileg%
\'{\i}timas e permite demonstrar que se o potencial singular for fracamente
atrativo n\~{a}o h\'{a} cabimento em se falar em colapso para o centro ou
inexist\^{e}ncia de estado fundamental. Finalmente, mostramos que, seja o
problema definido no semieixo ou em todo o eixo, o oscilador n\~{a}%
o-singular pode ser pensado como uma transi\c{c}\~{a}o de fase do oscilador
singular, e por causa disso a solu\c{c}\~{a}o do oscilador harm\^{o}nico
singular n\~{a}o pode ser obtida a partir da solu\c{c}\~{a}o do oscilador
harm\^{o}nico n\~{a}o-singular via teoria da perturba\c{c}\~{a}o.

\section{Potenciais singulares e degeneresc\^{e}ncia}

A equa\c{c}\~{a}o de Schr\"{o}dinger unidimensional para uma part\'{\i}cula
de massa de repouso $m$ sujeita a um potencial externo $V\left( x,t\right) $
\'{e} dada por%
\begin{equation}
i\hbar \,\frac{\partial \Psi }{\partial t}=\mathcal{H}\Psi ,  \label{1a}
\end{equation}%
Aqui, $\Psi \left( x,t\right) $ \'{e} a fun\c{c}\~{a}o de onda, $\hbar $
\'{e} a constante de Planck reduzida ($\hbar =h/(2\pi )$) e $\mathcal{H}$
\'{e} o operador hamiltoniano
\begin{equation}
\mathcal{H}=-\frac{\hbar ^{2}}{2m}\frac{\partial ^{2}}{\partial x^{2}}+V.
\label{1}
\end{equation}%
N\~{a}o \'{e} dificil mostrar que%
\begin{equation}
\frac{\partial |\Psi |^{2}}{\partial t}=\frac{i}{\hbar }\left[ \left(
\mathcal{H}\Psi \right) ^{\ast }\Psi -\Psi ^{\ast }\left( \mathcal{H}\Psi
\right) \right] ,
\end{equation}%
e levando em considera\c{c}\~{a}o que o operador hamiltoniano \'{e} um
operador hermitiano,\footnote{%
Todas as quantidades f\'{\i}sicas observ\'{a}veis correspondem a operadores
hermitianos. O operador $\mathcal{O}$ \'{e} dito ser hermitiano se%
\begin{equation*}
\int_{-\infty }^{+\infty }dx\,\left( \mathcal{O}\Psi _{1}\right) ^{\ast
}\Psi _{2}=\int_{-\infty }^{+\infty }dx\,\Psi _{1}^{\ast }\left( \mathcal{O}%
\Psi _{2}\right) ,
\end{equation*}%
onde $\Psi _{1}$ e $\Psi _{2}$ s\~{a}o duas fun\c{c}\~{o}es de onda
quaisquer que fazem $\int_{-\infty }^{+\infty }dx\,\Psi _{1}^{\ast }\left(
\mathcal{O}\Psi _{2}\right) <\infty $. Em particular, as fun\c{c}\~{o}es de
onda devem ser quadrado-integr\'{a}veis, viz. $\int_{-\infty }^{+\infty
}dx\,|\Psi |^{2}$ $<\infty $.} temos o corol\'{a}rio%
\begin{equation}
\frac{d}{dt}\int_{-\infty }^{+\infty }dx\,|\Psi |^{2}=0.
\end{equation}%
A equa\c{c}\~{a}o da continuidade%
\begin{equation}
\frac{\partial \rho }{\partial t}+\frac{\partial J}{\partial x}=0
\label{con}
\end{equation}%
\noindent \'{e} satisfeita com a densidade de probabilidade%
\begin{equation}
\rho =|\Psi |^{2}\quad  \label{rho}
\end{equation}%
e a corrente%
\begin{equation}
J=\frac{\hbar }{m}\,\textrm{Im}\left( \Psi ^{\ast }\frac{\partial \Psi }{%
\partial x}\right) .  \label{J}
\end{equation}%
A equa\c{c}\~{a}o da continuidade pode tamb\'{e}m ser escrita como%
\begin{equation}
J\left( x,t\right) -J\left( x_{0},t\right) =-\frac{d}{dt}\int_{x_{0}}^{x}d%
\eta \,\rho \left( \eta ,t\right) ,  \label{ec}
\end{equation}%
onde $x_{0}$ \'{e} um ponto arbitr\'{a}rio do eixo $X$. A forma integral da
equa\c{c}\~{a}o da continuidade, (\ref{ec}), permite interpretar
inequivocamente a corrente $J\left( x,t\right) $ como sendo o fluxo de
probabilidade atrav\'{e}s de $x$ no instante $t$.

No caso de potenciais externos independentes do tempo, a fun\c{c}\~{a}o de
onda $\Psi $ admite solu\c{c}\~{o}es particulares da forma

\begin{equation}
\Psi (x,t)=\psi (x)\,e^{-i\frac{E}{\hbar }t},  \label{2a}
\end{equation}

\noindent onde $\psi $ obedece \`{a} equa\c{c}\~{a}o de Schr\"{o}dinger
independente do tempo%
\begin{equation}
\mathcal{H}\psi =E\psi .  \label{auto}
\end{equation}%
Neste caso, com condi\c{c}\~{o}es de contorno apropriadas, o problema se
reduz \`{a} determina\c{c}\~{a}o do par caracter\'{\i}stico $\left( E,\psi
\right) $. A equa\c{c}\~{a}o de autovalor (\ref{auto}) tamb\'{e}m pode ser
escrita na forma

\begin{equation}
\frac{d^{2}\psi }{dx^{2}}+\left( k^{2}-\frac{2mV}{\hbar ^{2}}\right) \psi =0,
\label{3}
\end{equation}

\noindent com%
\begin{equation}
k^{2}=\frac{2mE}{\hbar ^{2}}.  \label{k2}
\end{equation}%
A densidade e a corrente correspondentes \`{a} solu\c{c}\~{a}o expressa por (%
\ref{2a}) tornam-se
\begin{equation}
\rho =\left\vert \psi \right\vert ^{2},\quad J=\frac{\hbar }{m}\,\textrm{Im}%
\left( \psi ^{\ast }\frac{d\psi }{dx}\right) .  \label{rho1}
\end{equation}%
Em virtude de $\rho $ e $J$ serem independentes do tempo, a solu\c{c}\~{a}o (%
\ref{2a}) \'{e} dita descrever um estado estacion\'{a}rio. Note que, por
causa da equa\c{c}\~{a}o da continuidade (\ref{con}) e (\ref{ec}), a
corrente $J$ n\~{a}o \'{e} t\~{a}o somente estacion\'{a}ria, mas tamb\'{e}m
uniforme, i.e. $J\left( x\right) =J\left( x_{0}\right) $.

Os estados ligados constituem uma classe de solu\c{c}\~{o}es da equa\c{c}%
\~{a}o de Schr\"{o}dinger que representam um sistema localizado numa regi%
\~{a}o finita do espa\c{c}o. Para estados ligados devemos procurar autofun%
\c{c}\~{o}es que se anulam \`{a} medida que $|x|\rightarrow \infty $. \'{E}
\'{o}bvio que, em decorr\^{e}ncia deste comportamento assint\'{o}tico, $%
J\rightarrow 0$ quando $|x|\rightarrow \infty $. Assim, a uniformidade da
corrente dos estados estacion\'{a}rios demanda que $J$ seja nula em todo o
espa\c{c}o. Fato esperado em vista da interpreta\c{c}\~{a}o de $J$
apresentada anteriormente. Tamb\'{e}m, neste caso podemos normalizar $\psi $
fazendo%
\begin{equation}
\int_{-\infty }^{+\infty }dx\,|\psi |^{2}=1.  \label{normaliza}
\end{equation}%
Com um potencial invariante sob reflex\~{a}o atrav\'{e}s da origem ($%
x\rightarrow -x$), autofun\c{c}\~{o}es com paridades bem definidas podem ser
constru\'{\i}das. Neste caso, as autofun\c{c}\~{o}es de $\mathcal{H}$ s\~{a}%
o tamb\'{e}m autofun\c{c}\~{o}es do operador paridade, viz.%
\begin{eqnarray}
\mathcal{H}\psi _{n}^{\left( p\right) } &=&E_{n}^{\left( p\right) }\psi
_{n}^{\left( p\right) }  \notag \\
&&  \label{comu} \\
\mathcal{P}\psi _{n}^{\left( p\right) } &=&p\psi _{n}^{\left( p\right) },
\notag
\end{eqnarray}%
onde $\mathcal{P}$ \'{e} o operador paridade, $p=\pm 1$ e $n$ denota
quaisquer outros n\'{u}meros qu\^{a}nticos. Como consequ\^{e}ncia da
hemiticidade de $\mathcal{H}$ e $\mathcal{P}$, as autofun\c{c}\~{o}es
satisfazem \`{a} condi\c{c}\~{a}o de ortogonalidade%
\begin{equation}
\int_{-\infty }^{+\infty }dx\,\left( \psi _{\tilde{n}}^{\left( \tilde{p}%
\right) }\right) ^{\ast }\psi _{n}^{\left( p\right) }=0,\quad \textrm{para }%
\tilde{n}\neq n\textrm{ ou }\tilde{p}\neq p.
\end{equation}%
Por causa da paridade, podemos concentrar a aten\c{c}\~{a}o no semieixo
positivo e impor condi\c{c}\~{o}es de contorno na origem e no infinito.
Naturalmente, a autofun\c{c}\~{a}o \'{e} cont\'{\i}nua. No entanto, temos de
considerar a condi\c{c}\~{a}o de conex\~{a}o entre a derivada primeira da
autofun\c{c}\~{a}o \`{a} direita e \`{a} esquerda da origem, que deve ser
obtida diretamente da equa\c{c}\~{a}o de Schr\"{o}dinger independente do
tempo. Normalizabilidade, conforme comentado de pouco, requer que $\psi
\rightarrow 0$ quando $|x|\rightarrow \infty $. Autofun\c{c}\~{o}es com
paridades bem definidas em todo o eixo podem ser constru\'{\i}das tomando
combina\c{c}\~{o}es lineares sim\'{e}tricas e antissim\'{e}tricas de \ $\psi
$ definida no lado positivo do eixo $X$. Estas novas autofun\c{c}\~{o}es
possuem a mesma energia, ent\~{a}o, em princ\'{\i}pio, existe uma dupla
degeneresc\^{e}ncia ($E_{n}^{\left( +\right) }=E_{n}^{\left( -\right) }$).
\'{E} not\'{o}rio que o espectro de estados ligados de sistemas
unidimensionais com potenciais regulares \'{e} n\~{a}o-degenerado (veja,
e.g., \cite{lan} e \cite{cm}). Entretanto, se o potencial for singular na
origem, por exemplo, tanto as autofun\c{c}\~{o}es pares quanto as autofun%
\c{c}\~{o}es \'{\i}mpares poderiam obedecer \`{a} condi\c{c}\~{a}o homog\^{e}%
nea de Dirichlet na origem, e cada n\'{\i}vel de energia exibiria uma
degeneresc\^{e}ncia de grau dois.\footnote{%
Os mais c\'{e}ticos quanto \`{a} possibilidade de degeneresc\^{e}ncia em
sistemas unidimensionais podem constatar esta particularidade visualizando
as poss\'{\i}veis autofun\c{c}\~{o}es no caso de dois po\c{c}os infinitos
dispostos simetricamente em torno de $x=0$.} A condi\c{c}\~{a}o de conex\~{a}%
o obedecida pela derivada primeira da autofun\c{c}\~{a}o, contudo, poderia
excluir uma das duas combina\c{c}\~{o}es lineares, e nesse caso os n\'{\i}%
veis de energia seriam n\~{a}o-degenerados.

\section{Oscilador singular}

Seguindo a nota\c{c}\~{a}o da Ref. \cite{pr}, vamos agora considerar o
potencial%
\begin{equation}
V\left( x\right) =\frac{1}{2}m\omega ^{2}x^{2}+\frac{\hbar ^{2}\alpha }{%
2mx^{2}},\quad \omega >0.  \label{pot}
\end{equation}%
O par\^{a}metro adimensional $\alpha $ caracteriza tr\^{e}s diferentes
perfis para o potencial, como est\'{a} ilustrado na Figura 1.
\begin{figure}[th]
\begin{center}
\includegraphics[width=10cm]{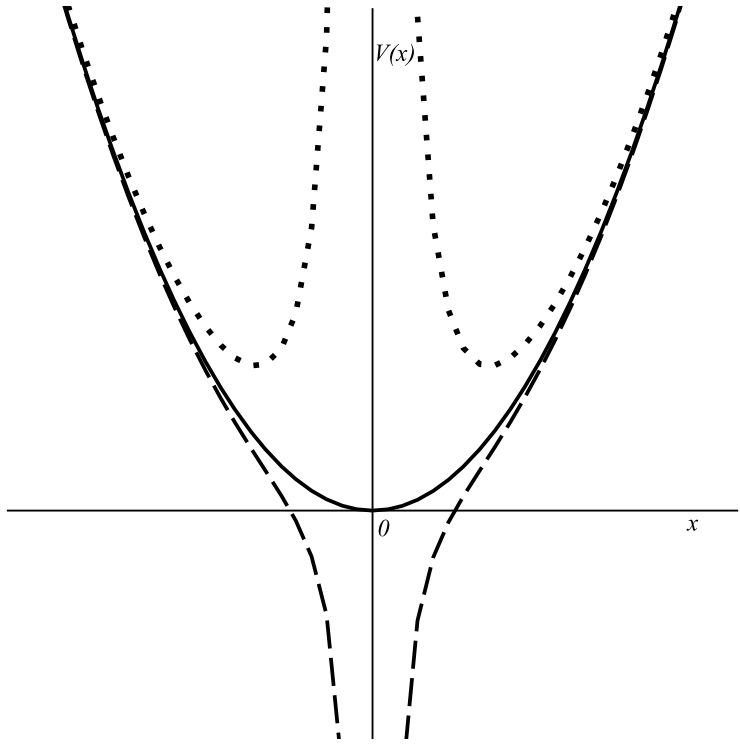} \label{fig:Fig1}
\end{center}
\par
\vspace*{-0.1cm}
\caption{Os tr\^{e}s perfis para $V\left( x\right) .$ As linhas tracejada,
cont\'{\i}nua e pontilhada para os casos com $\protect\alpha $ negativo,
nulo e positivo, respectivamente. }
\end{figure}
Para $\alpha =0$ temos o potencial do oscilador harm\^{o}nico regular (po%
\c{c}o simples), e para $\alpha \neq 0$ temos o caso de um po\c{c}o duplo
com uma barreira de potencial repulsiva e singular na origem ($\alpha >0$)
ou o caso de um po\c{c}o sem fundo puramente atrativo ($\alpha <0$). Para o
potencial (\ref{pot}), a equa\c{c}\~{a}o de Schr\"{o}dinger independente do
tempo assume a forma

\begin{equation}
\frac{d^{2}\psi }{dx^{2}}+\left( k^{2}-\lambda ^{2}x^{2}-\frac{\alpha }{x^{2}%
}\right) \psi =0,  \label{af}
\end{equation}%
com%
\begin{equation}
\lambda =\frac{m\omega }{\hbar }.  \label{l}
\end{equation}

Quando $\alpha =0$, as solu\c{c}\~{o}es da equa\c{c}\~{a}o (\ref{af}) s\~{a}%
o anal\'{\i}ticas em todo o eixo $X$. Quando $\alpha \neq 0$, todavia, $\psi
$ pode manifestar singularidade na origem. Tal singularidade poderia
comprometer a nulidade da corrente em $x=0$, a exist\^{e}ncia das integrais
definidas nos intervalos ($0,+\infty $) e ($-\infty ,+\infty $), e a
hermiticidade dos operadores associados com as quantidades f\'{\i}sicas
observ\'{a}veis. Por causa desta sensa\c{c}\~{a}o de amea\c{c}a, come\c{c}%
aremos a abordagem do problema pela averigua\c{c}\~{a}o do comportamento das
solu\c{c}\~{o}es de (\ref{af}) na vizinhan\c{c}a da origem. \'{E} claro que
o comportamento assint\'{o}tico ($|x|\rightarrow \infty $) tamb\'{e}m \'{e}
merit\'{o}rio.

\subsection{Comportamento na origem}

Na vizinhan\c{c}a da origem a equa\c{c}\~{a}o (\ref{af}) passa a ter duas
formas distintas:

\begin{eqnarray}
\frac{d^{2}\psi }{dx^{2}}+k^{2}\psi  &\simeq &0,\quad \textrm{para }\alpha =0
\notag \\
&&  \label{orig} \\
\frac{d^{2}\psi }{dx^{2}}-\frac{\alpha }{x^{2}}\psi  &\simeq &0,\quad \textrm{%
para }\alpha \neq 0.  \notag
\end{eqnarray}%
De um jeito ou de outro, no semieixo positivo podemos escrever%
\begin{equation}
\psi \simeq \left\{
\begin{array}{c}
A\,|x|^{\beta _{+}+1}+B\,|x|^{\beta _{-}+1},\quad \textrm{para }\alpha \neq
-1/4 \\
\\
D\,\,|x|^{1/2}+F\,\,|x|^{1/2}\log |x|,\quad \textrm{para }\alpha =-1/4,%
\end{array}%
\right.   \label{original}
\end{equation}%
onde%
\begin{equation}
\beta _{\pm }=-\frac{1}{2}\pm \sqrt{\frac{1}{4}+\alpha }  \label{Beta}
\end{equation}%
\'{e} solu\c{c}\~{a}o da equa\c{c}\~{a}o alg\'{e}brica indicial%
\begin{equation}
\beta _{\pm }\left( \beta _{\pm }+1\right) =\alpha .  \label{indi}
\end{equation}%
Na vizinhan\c{c}a da origem, o comportamento dos termos
\begin{eqnarray}
K_{\tilde{n}n} &=&\psi _{\tilde{n}}^{\ast }\left( -\frac{\hbar ^{2}}{2m}%
\frac{d^{2}}{dx^{2}}\right) \psi _{n}  \notag \\
&& \\
V_{\tilde{n}n} &=&\psi _{\tilde{n}}^{\ast }\left( \frac{\hbar ^{2}}{2m}\frac{%
\alpha }{x^{2}}\right) \psi _{n}  \notag
\end{eqnarray}%
comina a hermiticidade dos operadores associados com a energia cin\'{e}tica
e com o potencial. Para $\alpha \neq -1/4$, podemos escrever%
\begin{eqnarray}
K_{\tilde{n}n} &\simeq &-\frac{\hbar ^{2}\alpha }{2m}\left[ A_{\tilde{n}%
}^{\ast }A_{n}\,|x|^{2\textrm{{\small Re}}\beta _{+}}+B_{\tilde{n}}^{\ast
}B_{n}\,|x|^{2\textrm{{\small Re}}\beta _{-}}\right.   \notag \\
&& \\
&&+\left. A_{\tilde{n}}^{\ast }B_{n}\,|x|^{\beta _{+}^{\ast }+\beta
_{-}}+A_{n}B_{\tilde{n}}^{\ast }\,|x|^{\beta _{+}+\beta _{-}^{\ast }}\right]
,  \notag
\end{eqnarray}%
e para $\alpha =-1/4$%
\begin{eqnarray}
K_{\tilde{n}n} &\simeq &-\frac{\hbar ^{2}\alpha }{2m}\,\frac{1}{|x|}\,\left[
D_{\tilde{n}}^{\ast }D_{n}+F_{\tilde{n}}^{\ast }F_{n}\log ^{2}|x|\right.
\notag \\
&& \\
&&+\left. \left( D_{\tilde{n}}^{\ast }F_{n}+D_{n}F_{\tilde{n}}^{\ast
}\right) \log |x|\right] .  \notag
\end{eqnarray}%
Em ambos os casos $V_{\tilde{n}n}\simeq -K_{\tilde{n}n}$. Vemos destas \'{u}%
ltimas rela\c{c}\~{o}es que a hermiticidade do operador associado com a
energia cin\'{e}tica (ou potencial) \'{e} verificada somente se \textrm{%
Re\thinspace }$\beta _{\pm }>-1/2$ quando $\alpha \neq 0$, o que equivale a
dizer que o sinal negativo defronte do radical em (\ref{Beta}) deve ser
descartado e $\alpha $ deve ser maior que $-1/4$. Naturalmente, devemos
considerar $\beta _{-}=-1$ tanto quanto $\beta _{+}=0$ quando $\alpha =0$.
Portanto, podemos afirmar que $\psi $ comporta-se na vizinhan\c{c}a da
origem como%
\begin{equation}
Cx^{\beta +1}
\end{equation}%
com%
\begin{equation}
\beta =\left\{
\begin{array}{c}
-\frac{1}{2}+\sqrt{\frac{1}{4}+\alpha }, \\
\\
-1\textrm{ ou }0,%
\end{array}%
\begin{array}{c}
\quad \textrm{para }\alpha \neq 0\textrm{, }\quad \textrm{com }\alpha >-\frac{1}{4}
\\
\\
\textrm{para }\alpha =0.%
\end{array}%
\right. \textrm{ }
\end{equation}%
Note ainda que a nulidade da corrente requer apenas que $\beta \in
\mathbb{R}
$ e $F=0$, enquanto a condi\c{c}\~{a}o de ortonormalizabilidade exige apenas
que \textrm{Re}${\,}\beta _{\pm }>-3/2$ e $F=0$. A condi\c{c}\~{a}o $\alpha
>-1/4$, que nos faz evitar um potencial atrativo com singularidade muito
forte, relacionado com o problema da \textquotedblleft queda para o
centro\textquotedblright\ \cite{lan}, tanto quanto a solu\c{c}\~{a}o
apropriada da equa\c{c}\~{a}o indicial (\ref{indi}), foram obtidas aqui de
uma maneira extremamente simples, sem recorrer ao processo de regulariza\c{c}%
\~{a}o do potencial na origem.\footnote{%
No processo de regulariza\c{c}\~{a}o, $V(x)$ \'{e} substitu\'{\i}do por $%
V(x_{0})$ \ para $x<x_{0}\approx 0$ e depois de usar as condi\c{c}\~{o}es de
continuidade para $\psi $ e $d\psi /dx$ no \textit{cutoff} \ tomamos o
limite $x_{0}\rightarrow 0$. Resulta que a solu\c{c}\~{a}o com $\beta _{-}$
\'{e} suprimida em rela\c{c}\~{a}o a essa envolvendo $\beta _{+}$ quando $%
x_{0}\rightarrow 0$.} Nota-se que, ainda que acrescido da exig\^{e}ncia de
normalizabilidade, o processo de regulariza\c{c}\~{a}o do potencial \'{e}
inapto para exluir o caso com $\alpha \leq -1/4$ \cite{lan}. O crit\'{e}rio
de hermiticidade do operador associado com a energia cin\'{e}tica (ou
potencial) \'{e} l\'{\i}cito e suficiente para descartar solu\c{c}\~{o}es esp%
\'{u}rias.\footnote{%
A ortonormalizabilidade das autofun\c{c}\~{o}es (relacionada com a
hermiticidade do operador hamiltoniano) e a hermiticidade do operador
momento (relacionada com a nulidade da corrente) s\~{a}o crit\'{e}rios mais
fr\'{a}geis porque envolvem o comportamento de $\psi _{\tilde{n}}^{\ast
}\psi _{n}$ e $\psi _{\tilde{n}}^{\ast }d\psi _{n}/dx$ na vizinhan\c{c}a da
origem, respectivamente.} Resumidamente,%
\begin{equation}
\beta =-1\textrm{ ou }\beta >-\frac{1}{2}.  \label{beta2}
\end{equation}%
A condi\c{c}\~{a}o de Dirichlet homog\^{e}nea ($\psi \left( 0\right) =0$)
\'{e} essencial sempre que $\alpha \neq 0$, contudo ela tamb\'{e}m ocorre
para $\alpha =0$ quando $\beta =0$ mas n\~{a}o para $\beta =-1$. Em suma,%
\begin{equation}
\left. \psi \right\vert _{x=0_{+}}\simeq \left\{
\begin{array}{c}
0, \\
\\
C,%
\end{array}%
\begin{array}{c}
\quad \textrm{para }\alpha \neq 0\textrm{, }\quad \textrm{ou }\alpha =0\textrm{ e }%
\beta =0 \\
\\
\quad \textrm{para }\alpha =0\textrm{ e }\beta =-1,%
\end{array}%
\right.
\end{equation}%
e%
\begin{equation}
\left. \frac{d\psi }{dx}\right\vert _{x=0_{+}}\simeq \left\{
\begin{array}{c}
0, \\
\\
C, \\
\\
\infty ,%
\end{array}%
\begin{array}{c}
\quad \textrm{para }\alpha >0\textrm{,}\quad \textrm{ou }\alpha =0\textrm{ e }\beta
=-1 \\
\\
\quad \textrm{para }\alpha =0\textrm{ e }\beta =0 \\
\\
\quad \textrm{para }\alpha <0.%
\end{array}%
\right.
\end{equation}

\subsection{Comportamento assint\'{o}tico}

A equa\c{c}\~{a}o de Schr\"{o}dinger independente do tempo para o nosso
problema, Eq. (\ref{af}), tem o comportamento assint\'{o}tico ($%
|x|\rightarrow \infty $)
\begin{equation}
\frac{d^{2}\psi }{dx^{2}}-\lambda ^{2}x^{2}\psi \simeq 0,
\end{equation}%
e da\'{\i} sucede que a forma assint\'{o}tica da solu\c{c}\~{a}o
quadrado-integr\'{a}vel \'{e} dada por
\begin{equation}
\psi \simeq e^{-\lambda x^{2}/2}.  \label{asym}
\end{equation}

\subsection{Solu\c{c}\~{a}o no semieixo}

O comportamento assint\'{o}tico de $\psi $ expresso por (\ref{asym})
convida-nos a definir $y=\lambda x^{2}$ de forma que a autofun\c{c}\~{a}o
para todo $y$ pode ser escrita como%
\begin{equation}
\psi \left( y\right) =y^{\left( \beta +1\right) /2}\,e^{-y/2}\,w\left(
y\right) ,  \label{eq4}
\end{equation}%
onde a fun\c{c}\~{a}o desconhecida $w\left( y\right) $ \'{e} solu\c{c}\~{a}o
regular da equa\c{c}\~{a}o hipergeom\'{e}trica confluente \cite{abr}
\begin{equation}
y\,\frac{d^{\,2}w\left( y\right) }{dy^{2}}+(b-y)\,\frac{dw\left( y\right) }{%
dy}-a\,w\left( y\right) =0,  \label{kum}
\end{equation}%
\noindent com%
\begin{equation}
a=\frac{b}{2}-\frac{k^{2}}{4\lambda }\quad \textrm{e}\quad b=\beta +\frac{3}{2}%
.  \label{ab}
\end{equation}%
A solu\c{c}\~{a}o geral de (\ref{kum}) \'{e} dada por \cite{abr}
\begin{equation}
w\left( y\right) =A\,M(a,b,y)+B\,y^{1-b}\,M(a-b+1,2-b,y),  \label{sg}
\end{equation}%
onde$\ A$ e $B$ s\~{a}o constantes arbitr\'{a}rias, e $M(a,b,y)$, tamb\'{e}m
denotada por $_{1}F_{1}\left( a,b,y\right) $, \'{e} a fun\c{c}\~{a}o
hipergeom\'{e}trica confluente (tamb\'{e}m chamada de fun\c{c}\~{a}o de
Kummer) expressa pela s\'{e}rie \cite{abr}%
\begin{equation}
M(a,b,y)=\frac{\Gamma \left( b\right) }{\Gamma \left( a\right) }%
\sum_{j=0}^{\infty }\frac{\Gamma \left( a+j\right) }{\Gamma \left(
b+j\right) }\,\frac{y^{j}}{j!},  \label{ser}
\end{equation}%
onde $\Gamma \left( z\right) $ \'{e} a fun\c{c}\~{a}o gama. \noindent A fun%
\c{c}\~{a}o gama n\~{a}o tem ra\'{\i}zes e seus polos s\~{a}o dados por $z=-n
$, onde $n$ \'{e} um inteiro n\~{a}o-negativo \cite{abr}. A fun\c{c}\~{a}o
de Kummer converge para todo $y$, \'{e} regular na origem ($M(a,b,0)=1$) e
tem o comportamento assint\'{o}tico prescrito por \cite{abr}%
\begin{equation}
M(a,b,y)\,_{\simeq }\,\frac{\Gamma \left( b\right) }{\Gamma \left(
b-a\right) }\,e^{-i\pi a}\,y^{-a}+\frac{\Gamma \left( b\right) }{\Gamma
\left( a\right) }\,e^{y}\,y^{a-b}.  \label{asy}
\end{equation}%
Haja vista que $b>1$, e estamos em busca de solu\c{c}\~{a}o regular na
origem, devemos tomar%
\begin{equation}
B=0
\end{equation}%
em (\ref{sg}). A presen\c{c}a de $e^{y}$ em (\ref{asy}) deprava o bom
comportamento assint\'{o}tico da autofun\c{c}\~{a}o j\'{a} ditado por (\ref%
{asym}). Para remediar esta situa\c{c}\~{a}o constrangedora devemos
considerar os polos de $\Gamma \left( a\right) $, e assim preceituar que um
comportamento aceit\'{a}vel para $M(a,b,y)$ ocorre somente se%
\begin{equation}
a=-n,\quad n\in
\mathbb{N}
.  \label{n}
\end{equation}%
Neste caso, a s\'{e}\-rie (\ref{ser}) \'{e} truncada em $j=n$ e o polin\^{o}%
mio de grau $n$ resultante \'{e} proporcional ao polin\^{o}mio de Laguerre
generalizado $L_{n}^{\left( b-1\right) }\left( y\right) $, com $b>0$ \cite%
{abr}. Portanto, de (\ref{k2}), (\ref{l}), (\ref{ab}) e (\ref{n}) podemos
determinar que as energias permitidas s\~{a}o dadas por
\begin{equation}
E_{n}=\left( 2n+\beta +\frac{3}{2}\right) \hbar \omega ,  \label{energia}
\end{equation}%
e as autofun\c{c}\~{o}es definidas no semieixo positivo s\~{a}o%
\begin{equation}
\psi _{n}\left( |x|\right) =A_{n}\,|x|^{\beta +1}\,e^{-\lambda
x^{2}/2}\,L_{n}^{\left( \beta +1/2\right) }(\lambda x^{2}).  \label{afu}
\end{equation}

Na Figura 2
\begin{figure}[th]
\begin{center}
\includegraphics[width=10cm]{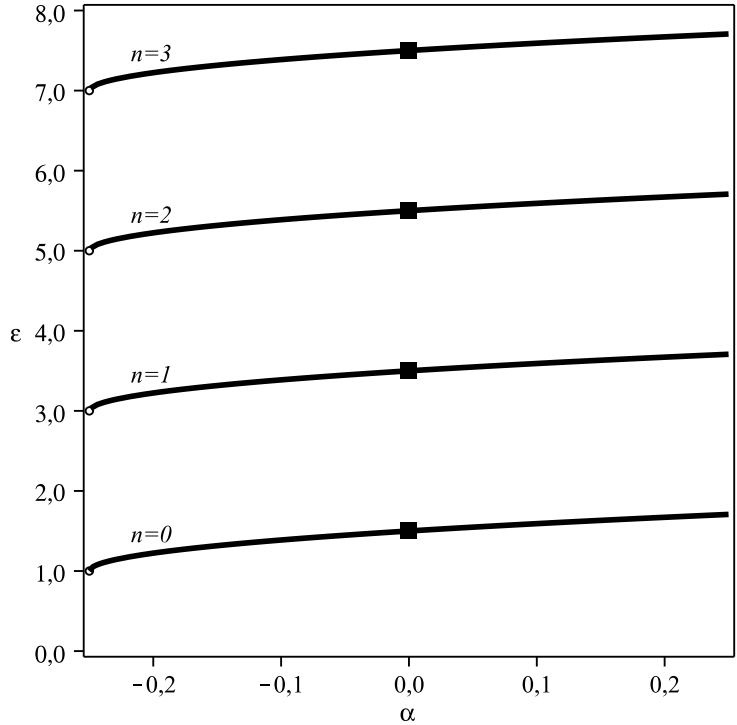} \label{fig:Fig2}
\end{center}
\par
\vspace*{-0.1cm}
\caption{Autoenergias ($\protect\epsilon =E/(\hbar \protect\omega)$) em fun%
\c{c}\~{a}o de $\protect\alpha$ no intervalo $(-1/4,+1/4]$. As linhas cont%
\'{\i}nuas para o oscilador singular, e os quadrados em $\protect\alpha =0$
para o oscilador n\~{a}o-singular.}
\end{figure}
ilustramos os primeiros n\'{\i}veis de energia em fun\c{c}\~{a}o de $\alpha $
no intervalo $(-1/4,+1/4]$. Observe que, qualquer que seja $\alpha $, ainda
que $-1/4<\alpha <0$, o espectro \'{e} discreto e sempre positivo. H\'{a} um
n\'{u}mero infinito de n\'{\i}veis de energia igualmente espa\c{c}ados (espa%
\c{c}amento igual a $2\hbar \omega $, independentemente de $\alpha $), e o
estado fundamental tem energia $\left( 2\beta +3\right) \hbar \omega
/2>\hbar \omega $.

Na Figura 3
\begin{figure}[th]
\begin{center}
\includegraphics[width=10cm]{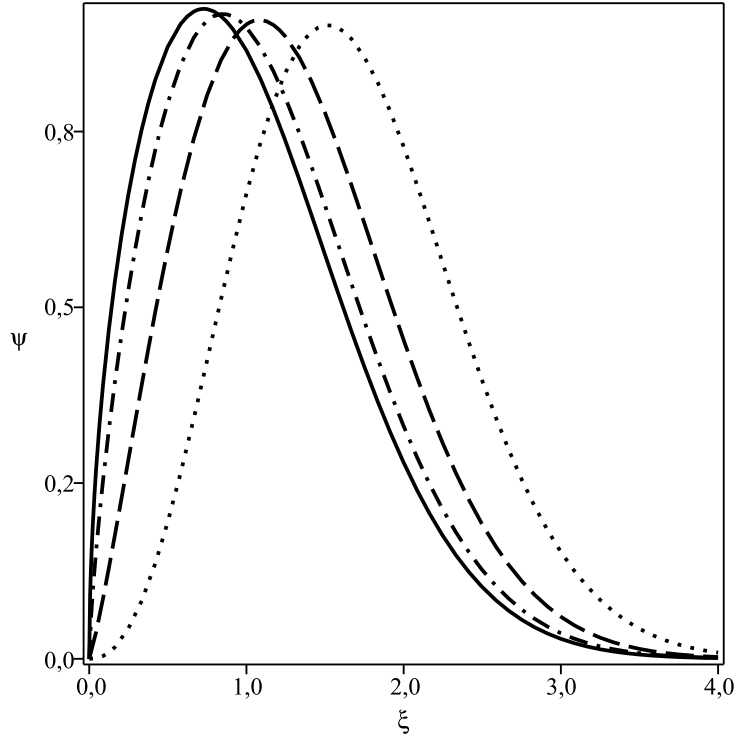} \label{fig:Fig3}
\end{center}
\par
\vspace*{-0.1cm}
\caption{Autofun\c{c}\~{a}o normalizada do estado fundamental definido no
semieixo em fun\c{c}\~{a}o de $\protect\xi =\protect\sqrt{\protect\lambda }{%
\,}x$. As linhas cont\'{\i}nua, ponto-tracejada, tracejada e pontilhada para
os casos com $\protect\alpha $ igual a $-0,249$, $-0,2$, $+0,2$ e $+3$,
respectivamente.}
\end{figure}
ilustramos o comportamento da autofun\c{c}\~{a}o para o estado fundamental.
A normaliza\c{c}\~{a}o foi realizada por m\'{e}todos num\'{e}ricos mas
poderia ter sido obtida por meio de f\'{o}rmulas envolvendo os polin\^{o}%
mios de Laguerre associados constantes na Ref. \cite{abr}. A compara\c{c}%
\~{a}o entre as quatro curvas mostra que a part\'{\i}cula tende a evitar a
origem mais e mais \`{a} medida que $\alpha $ aumenta. Eis um resultado
esperado no caso $\alpha >0$ que ocorre tamb\'{e}m no caso $\alpha <0$.
Observe que, qualquer que seja $\alpha >-1/4$, as autofun\c{c}\~{o}es (\ref%
{afu}) s\~{a}o fisicamente aceit\'{a}veis, ainda que no intervalo$\
-1/4<\alpha <0$ elas possuam derivada primeira singular. \'{E} mesmo assim,
contanto que o par\^{a}metro $\alpha $ seja maior que $-1/4$, o par caracter%
\'{\i}stico $\left( E_{n},\psi _{n}\right) $ constitui uma solu\c{c}\~{a}o
permiss\'{\i}vel do problema proposto. A part\'{\i}cula nunca colapsa para o
ponto $x=0$ e certamente h\'{a} um estado fundamental.

\subsection{Solu\c{c}\~{a}o em todo o eixo}

A autofun\c{c}\~{a}o definida para todo o eixo $X$ pode ser escrita como%
\begin{equation}
\psi _{n}^{\left( p\right) }\left( x\right) =\theta \left( x\right) \psi
_{n}\left( |x|\right) +p\,\theta \left( -x\right) \psi _{n}\left( |x|\right)
,  \label{todoeixo}
\end{equation}%
onde%
\begin{equation}
\theta \left( x\right) =\left\{
\begin{array}{cc}
1 & {\textrm{para }}x>0, \\
&  \\
0 & \textrm{para }x<0%
\end{array}%
\right.  \label{l5}
\end{equation}%
\'{e} a fun\c{c}\~{a}o degrau de Heaviside, $p$ \'{e} o autovalor do
operador paridade, e a autoenergia \'{e} dada por%
\begin{equation}
E_{n}^{\left( p\right) }=\left( 2n+\beta +\frac{3}{2}\right) \hbar \omega .
\end{equation}%
A hermiticidade do operador associado com a energia cin\'{e}tica (ou
potencial), por causa da singularidade em $x=0$ no caso $\alpha <0$ ($\beta
<0$), depende da exist\^{e}ncia do valor principal de Cauchy\footnote{%
Se $f\left( x\right) $ for singular na origem, a integral $\int_{-\infty
}^{+\infty }dx\,\,f\left( x\right) $ ser\'{a} \textit{nonsense}. Contudo, o
valor principal de Cauchy, $P\int_{-\infty }^{+\infty }dx\,\,f\left(
x\right) $, \'{e} uma prescri\c{c}\~{a}o que pode atribuir um sentido
proveitoso \`{a} representa\c{c}\~{a}o integral por meio da receita que se
segue:
\begin{equation}
P\int_{-\infty }^{+\infty }dx\,\,f\left( x\right) =\underset{\varepsilon
\rightarrow 0}{\,\lim }\left( \int_{-\infty }^{-\varepsilon }dx\,\,f\left(
x\right) +\int_{+\varepsilon }^{+\infty }dx\,f\left( x\right) \right) .
\label{ecauchy}
\end{equation}%
} da integral%
\begin{equation}
\int_{-\infty }^{+\infty }dx\,K_{\tilde{n}n}^{\left( \tilde{p}p\right) }.
\label{intk}
\end{equation}%
Obviamente, o valor principal de Cauchy poderia consentir um afrouxamento
das condi\c{c}\~{o}es de contorno impostas sobre as autofun\c{c}\~{o}es.
Autofun\c{c}\~{o}es mais singulares que essas anteriormente definidas no
semieixo seriam toleradas se na vizinhan\c{c}a da origem os sinais de $K_{%
\tilde{n}n}^{\left( \tilde{p}p\right) }$ \`{a} direita e \`{a} esquerda da
origem fossem diferentes para quaisquer $p$ e $\tilde{p}$. Por\'{e}m, temos%
\begin{equation}
K_{\tilde{n}n}^{\left( \tilde{p}p\right) }\left( x<0\right) =\tilde{p}p\,K_{%
\tilde{n}n}\left( x>0\right)
\end{equation}%
de modo que, no caso em que $\alpha <0$, a integral (\ref{intk}) n\~{a}o
seria finita para $\tilde{p}=p$. Somos assim conduzidos a preservar a
rigidez do crit\'{e}rio de hermiticidade j\'{a} estabelecido no problema
definido no semieixo.

A continuidade (ou descontinuidade) de $d\psi /dx$ na origem pode ser
avaliada pela integra\c{c}\~{a}o de (\ref{orig}) de $-\varepsilon $ para $%
+\varepsilon $ no limite $\varepsilon \rightarrow 0$. A f\'{o}rmula de conex%
\~{a}o entre $d\psi /dx$ \`{a} direita e $d\psi /dx$ \`{a} esquerda da
origem pode ser sumarizada por%
\begin{equation}
\lim_{\varepsilon \rightarrow 0}\left. \frac{d\psi }{dx}\right\vert
_{x=-\varepsilon }^{x=+\varepsilon }=\alpha \lim_{\varepsilon
\longrightarrow 0}\int_{-\varepsilon }^{+\varepsilon }dx\;\frac{\psi }{x^{2}}%
.
\end{equation}%
Tomando em considera\c{c}\~{a}o o valor principal de Cauchy, no caso de $%
\psi $ antissim\'{e}trica com $\beta \neq 0$, podemos afirmar que as autofun%
\c{c}\~{o}es t\^{e}m derivada primeira cont\'{\i}nua na origem. Assim, o
espectro do oscilador singular \'{e} duplamente degenerado.

Por causa da continuidade, n\~{a}o h\'{a} autofun\c{c}\~{a}o \'{\i}mpar para
$\beta =-1$ e l\'{a} ocorre a condi\c{c}\~{a}o homog\^{e}nea de Neumann ($%
\left. d\psi /dx\right\vert _{x=0_{+}}=0$), e n\~{a}o h\'{a} autofun\c{c}%
\~{a}o par para $\beta =0$. Portanto, o espectro do oscilador regular \'{e} n%
\~{a}o-degenerado, como deveria ser. Para $\beta =-1$ e $\beta =0$, em
particular, os polin\^{o}mios de Laguerre $L_{n}^{\left( -1/2\right)
}(\lambda x^{2})$ e $L_{n}^{\left( +1/2\right) }(\lambda x^{2})$ s\~{a}o
proporcionais aos polin\^{o}mios de Hermite $H_{2n}\left( \sqrt{\lambda }%
|x|\right) $ e $x^{-1}H_{2n+1}\left( \sqrt{\lambda }|x|\right) $,
respectivamente \cite{abr}:%
\begin{eqnarray}
L_{n}^{\left( -1/2\right) }(\lambda x^{2}) &=&\frac{\left( -1\right) ^{n}}{%
n!\,2^{2n}}\,H_{2n}\left( \sqrt{\lambda }|x|\right)   \notag \\
&&  \label{her} \\
L_{n}^{\left( +1/2\right) }(\lambda x^{2}) &=&\frac{\left( -1\right) ^{n}}{%
n!\,2^{2n+1}\sqrt{\lambda }\,x}\,H_{2n+1}\left( \sqrt{\lambda }|x|\right) .
\notag
\end{eqnarray}%
Os polin\^{o}mios de Hermite s\~{a}o definidos no intervalo $\left( -\infty
,+\infty \right) $ e gozam da propriedade $H_{n}\left( -x\right) =\left(
-1\right) ^{n}H_{n}\left( x\right) $. Assim, a solu\c{c}\~{a}o do oscilador n%
\~{a}o-singular pode ser escrita na forma que se costumou em termos dos polin%
\^{o}mios de Hermite:%
\begin{equation}
E_{n}=\left( n+\frac{1}{2}\right) \hbar \omega ,
\end{equation}%
com autofun\c{c}\~{a}o definida em todo o eixo expressa por%
\begin{equation}
\psi _{n}\left( x\right) =A_{n}\,\,e^{-\lambda x^{2}/2}\,H_{n}\left( \sqrt{%
\lambda }x\right) .  \label{psi2}
\end{equation}%
As constantes $A_{n}$ em (\ref{afu}) e (\ref{psi2}), chamadas de constantes
de normaliza\c{c}\~{a}o, podem ser determinadas por meio da condi\c{c}\~{a}o
de normaliza\c{c}\~{a}o expressa por (\ref{normaliza}).

Na Figura 4
\begin{figure}[th]
\begin{center}
\includegraphics[width=10cm]{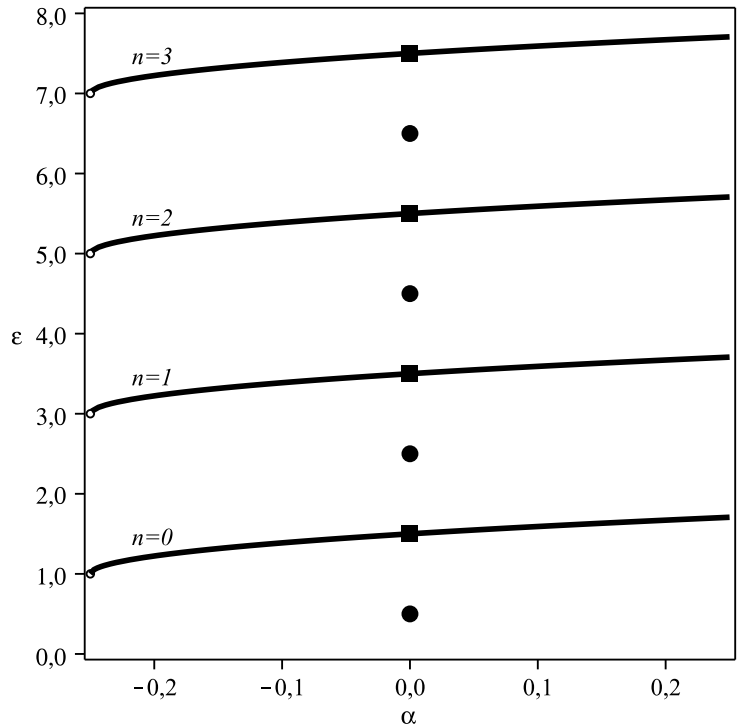} \label{fig:Fig4}
\end{center}
\par
\vspace*{-0.1cm}
\caption{Autoenergias ($\protect\epsilon =E/(\hbar \protect\omega)$) em fun%
\c{c}\~{a}o de $\protect\alpha$ no intervalo $(-1/4,+1/4]$. As linhas cont%
\'{\i}nuas para o oscilador singular, os quadrados e os c\'{\i}rculos em $%
\protect\alpha =0$ para as solu\c{c}\~{o}es \'{\i}mpares e pares do
oscilador harm\^{o}nico n\~{a}o-singular, respectivamente.}
\end{figure}
ilustramos os primeiros n\'{\i}veis de energia em fun\c{c}\~{a}o de $\alpha $
no intervalo $(-1/4,+1/4]$ para o problema definido em todo o eixo. Para $%
\alpha \neq 0$, o espectro \'{e} exatamente igual a esse do problema
definido no semieixo. Quando a singularidade \'{e} nula, entanto, o espa\c{c}%
amento dos n\'{\i}veis \'{e} $\hbar \omega $. Esta mudan\c{c}a de espa\c{c}%
amento de n\'{\i}veis quando $\alpha =0$ \'{e} devida aos n\'{\i}veis
intrusos que surgem por causa emerg\^{e}ncia da condi\c{c}\~{a}o de contorno
homog\^{e}nea de Neumann, em adi\c{c}\~{a}o \`{a} condi\c{c}\~{a}o de
contorno homog\^{e}nea de Dirichlet j\'{a} existente no problema definido no
semieixo. Esta invas\~{a}o de novas solu\c{c}\~{o}es com ${\psi }\left(
0\right) \neq 0$ tem como consequ\^{e}ncia imediata um dr\'{a}stico efeito
sobre a localiza\c{c}\~{a}o da part\'{\i}cula.

\section{Coment\'{a}rios finais}

Os resultados apresentados neste trabalho mostram com transpar\^{e}ncia que
o oscilador harm\^{o}nico singular unidimensional, seja definido no semieixo
ou em todo o eixo, seja repulsivo ou atrativo, exibe um conjunto infinito de
solu\c{c}\~{o}es aceit\'{a}veis, em claro contraste com os ditames
estampados nas refer\^{e}ncias \cite{bg} e \cite{pr}, e transcritos na
Introdu\c{c}\~{a}o. Sim, para um potencial singular fracamente atrativo o
colapso para o centro n\~{a}o tem nada a ver, e certamente h\'{a} um estado
fundamental com energia maior que $\hbar \omega /2$. A generaliza\c{c}\~{a}o
dos resultados do problema definido no semieixo para o oscilador harm\^{o}%
nico singular tridimensional pode ser feita com facilidade por meio da
substitui\c{c}\~{a}o de $\alpha $ por $\alpha +l\left( l+1\right) $, onde $l$
\'{e} o numero qu\^{a}ntico orbital, e pela concomitante substitui\c{c}\~{a}%
o de $\psi $ pela fun\c{c}\~{a}o radial $xR\left( x\right) $.

Visto como fun\c{c}\~{a}o de $\alpha $, o problema do oscilador singular
apresenta uma not\'{o}ria transi\c{c}\~{a}o de fase em $\alpha =0$.

Para o problema definido no semieixo, a transi\c{c}\~{a}o de fase
manifesta-se apenas por meio do comportamento da derivada primeira da autofun%
\c{c}\~{a}o na origem. Derivada primeira infinita para $\alpha <0$,
constante para $\alpha =0$, e nula para $\alpha >0$.

Para o problema definido em todo o eixo, entretanto, a transi\c{c}\~{a}o de
fase manifesta-se por meio da degeneresc\^{e}ncia, pelo comportamento da
autofun\c{c}\~{a}o e sua derivada primeira na origem, e pela localiza\c{c}%
\~{a}o da part\'{\i}cula. Uma outra assinatura da transi\c{c}\~{a}o de fase
\'{e} o espa\c{c}amento dos n\'{\i}veis de energia. Quando o potencial \'{e}
singular na origem, as autoenergias s\~{a}o igualmente espa\c{c}adas com
passo igual a $2\hbar \omega $. \'{E} admir\'{a}vel que esse espa\c{c}amento
de n\'{\i}veis \'{e} independente da intensidade do par\^{a}metro respons%
\'{a}vel pela singularidade do potencial. Quando a singularidade \'{e} nula,
entanto, o espa\c{c}amento dos n\'{\i}veis \'{e} $\hbar \omega $. Esta
brusca mudan\c{c}a de espa\c{c}amento de n\'{\i}veis quando $\alpha $ passa
por $\alpha =0$ \'{e} devida aos n\'{\i}veis intrusos que surgem por causa
emerg\^{e}ncia da condi\c{c}\~{a}o de contorno homog\^{e}nea de Neumann, em
adi\c{c}\~{a}o \`{a} condi\c{c}\~{a}o de contorno homog\^{e}nea de Dirichlet
j\'{a} existente para $\alpha \neq 0$. Esta intrus\~{a}o permite o
surgimento de polin\^{o}mios de Hermite pares e seus autovalores associados,
que se entremeiam entre os autovalores pr\'{e}-existentes associados com os
polin\^{o}mios de Hermite \'{\i}mpares. Os polin\^{o}mios de Hermite pares t%
\^{e}m ${\psi }_{n}^{\left( +\right) }\left( 0\right) \neq 0$ e esta condi%
\c{c}\~{a}o de contorno nunca \'{e} permitida quando a singularidade est\'{a}
presente, ainda que $\alpha $ seja muito pequeno. Esta invas\~{a}o s\'{u}%
bita dos polin\^{o}mios pares tem um brusco efeito sobre a localiza\c{c}\~{a}%
o da part\'{\i}cula. Poder-se-ia tamb\'{e}m tentar compreender tal transi%
\c{c}\~{a}o s\'{u}bita partindo de um potencial n\~{a}o-singular ($\alpha =0$%
), quando a solu\c{c}\~{a}o do problema envolve os polin\^{o}mios de Hermite
pares e \'{\i}mpares, e ent\~{a}o adicionar o potencial singular como uma
perturba\c{c}\~{a}o do potencial com $\alpha =0$. Agora, por sua natureza o
\textquotedblleft potencial singular perturbativo\textquotedblright\
repulsivo, demanda que $\psi _{n}^{\left( \pm \right) }\left( 0\right) =0$ e
assim ele \textquotedblleft mata\textquotedblright\ naturalmente a solu\c{c}%
\~{a}o envolvendo os polin\^{o}mios de Hermite pares. Ademais, n\~{a}o h\'{a}
degeneresc\^{e}ncia no espectro para o caso de $\alpha =0$.

Lathouwers \cite{lat} considerou o caso unidimensional do oscilador harm\^{o}%
nico singular como o oscilador harm\^{o}nico n\~{a}o-singular perturbado.
Acontece que, por causa dos distintos comportamentos da derivada primeira da
autofun\c{c}\~{a}o na origem para $\alpha =0$ ($\left. d\psi /dx\right\vert
_{x=0}=C$) e $\alpha \neq 0$ ($\left. d\psi /dx\right\vert _{x=0}=\infty $
para $\alpha <0$, e $\left. d\psi /dx\right\vert _{x=0}=0$ para $\alpha >0$%
), nossos coment\'{a}rios finais desfavorecem tal aspira\c{c}\~{a}o, ainda
que haja continuidade do espectro na vizinhan\c{c}a de $\alpha =0$ no caso
associado com autofun\c{c}\~{o}es \'{\i}mpares do oscilador n\~{a}o-singular.

\vspace{3cm}

\noindent{\textbf{Agradecimentos}}

O autores s\~{a}o gratos \`{a} FAPESP, ao CNPq e \`{a} CAPES pelo apoio financeiro. Um
\'{a}rbitro atento e ponderado contribuiu significantemente para a depura%
\c{c}\~{a}o e aprimoramento deste trabalho.

\end{document}